\documentstyle[aps,preprint]{revtex}

\begin{document}
\draft
\preprint{Fermilab-Pub-96/035-A, astro-ph/9603034}
\date{\today}
\title{The cosmological constant and \\
advanced gravitational wave detectors}

\author{Yun Wang$^{1}$ and Edwin L. Turner$^{2}$}
\address{
$^{1}$NASA/Fermilab Astrophysics Center, FNAL, Batavia, IL~~60510}

\address{
$^{2}$ Princeton University Observatory, Peyton Hall, Princeton, NJ~~08544}

\maketitle

\begin{abstract}

Interferometric gravitational wave detectors could measure the frequency 
sweep of a binary inspiral [characterized by its chirp mass] to high accuracy.
The observed chirp mass is the intrinsic chirp mass of the binary source
multiplied by $(1+z)$, where $z$ is the redshift of the source.
Assuming a non-zero cosmological constant,
we compute the expected redshift distribution of observed events for an 
advanced LIGO detector.
We find that the redshift distribution has a robust and sizable dependence 
on the cosmological constant; the data from advanced LIGO detectors
could provide an independent measurement of the cosmological constant.

\end{abstract}

\pacs{PACS numbers: 98.80.Es, 04.80.Nn }

\newpage

\narrowtext

\section{Introduction}

A non-zero cosmological constant may help solve some of the current
observational puzzles, most notably, the conflict between the 
age of globular clusters and the apparent
high value of the Hubble constant [which suggests a younger Universe].
\cite{Lambda}
A sizable cosmological constant
can make the Universe old in spite of a high Hubble constant,
although a non-zero cosmological constant is ugly from the
theoretical viewpoint. Whatever our aesthetic preferences, the value of
the cosmological constant should ultimately be determined by 
observational measurements.

Advanced LIGO detectors can expect to observe
approximately 50 neutron star binary inspiral events
per year, from distances up to 2000$\,$Mpc, the accuracy in
the measurement of the signal strength can be better than 10\%, and the
accuracy in the measurement of the chirp mass [which characterizes the 
frequency sweep of a binary inspiral] can be better than 0.1\%.
\cite{FinnCher93}\cite{CutFla94}.
The cosmological implications of gravitational wave observations of
binary inspiral have been discussed by several authors 
\cite{Schutz86,CherFinn93,Marko93}.
Most recently, Finn pointed out that the observations of 
binary inspirals in an interferometric gravitational wave detector,
in terms of the distribution of observed events with signal strength and chirp 
mass, can be quite sensitive to cosmology \cite{Finn96}.

Previous discussions of the cosmological implications of gravitational 
wave observations have considered the measurements of the
Hubble constant $H_0$ [$H_0=100h\,$km$\,$sec$^{-1}$Mpc$^{-1}$,
$0.5 \leq h <1$] and the deceleration parameter $q_0$,
assuming the cosmological constant to be zero.
Even though Markovi\'{c} discussed the measurement of the 
cosmological constant, he assumed its true value to be {\em zero}. \cite{Marko93}
In this paper, we consider the measurement of a {\it non-zero}
cosmological constant $\Omega_{\Lambda}$, and find that the 
$\Omega_{\Lambda}$ dependence of the observed chirp mass spectrum is
more robust than its $H_0$ dependence.

Markovi\'{c} proposed measuring the cosmological parameters
using the observed distribution of measured luminosity distances 
and (estimated) redshifts \cite{Marko93}, while Chernoff and Finn
suggested an alternative method without using the measured luminosity
distances \cite{CherFinn93}. We have chosen to follow the method 
and notations of Refs.{\cite{CherFinn93,Finn96}}.

\section{The chirp mass and the signal-to-noise ratio}

The observed chirp mass ${\cal M} \equiv {\cal M}_0 (1+z)$,
where ${\cal M}_0$ is the intrinsic chirp mass of the binary source,
and $z$ is the redshift of the source. Assuming a known and constant
${\cal M}_0$, the chirp mass spectrum is determined by
the source redshift distribution. For simplicity, we only
consider neutron star binaries in this paper;
for typical neutron star masses, ${\cal M}_0=1.19\,{\rm M}_{\odot}$. 
\cite{Finn96}
We compute the expected redshift distribution of 
neutron star binary events for an advanced LIGO detector, 
assuming a non-zero cosmological constant.

Neutron star binaries (NS/NS) may one day become the ``bread
and butter'' sources of LIGO style gravitational wave detectors.
NS/NS merger rate at redshift $z$ per unit observer time interval per unit volume is 
\begin{equation}
\frac{{\rm d} n}{{\rm d} t} = \dot{n}_0 \, (1+z)^2,
\end{equation}
where $\dot{n}_0$ is the local NS/NS merger rate per unit volume,
$(1+z)^2$ accounts for the shrinking of volumes with redshift (assuming 
constant comoving volume density of the merger rate)
and the time dilation.

The ``best guess'' local rate density from Phinney
\cite{Phinney91} is
\begin{equation}
\dot{n}_0 \simeq \left(9.9+ 0.6\, h^2\right)\, h \times 10^{-8} {\rm Mpc}^{-3} 
{\rm yr}^{-1} \simeq 10^{-7} h\, {\rm Mpc}^{-3} {\rm yr}^{-1}.
\label{eq:localRate}
\end{equation}
i.e., 3 per year at 200Mpc for $h=0.75$.
Ref.\cite{NaraPirShe91} gives $\dot{n}_0 \simeq 3h^3 \times 10^{-8} 
{\rm Mpc}^{-3}{\rm yr}^{-1}$, which is consistent with Eq.(\ref{eq:localRate})
within the uncertainty of the estimates.
Since the rate of star formation in galaxies appears to increase rapidly as
one looks back to $z=0.3$, it is possible that 
we underestimate the rates for LIGOs sensitive to sources
at cosmological distances unless we consider evolutionary effects. 
\cite{Phinney91}
In this paper, we neglect evolutionary effects.

LIGO aims to monitor the last stage of inspiral of a NS/NS binary,
during which the gravity waves generated sweep up in frequency,
over a time of about 15 minutes, from 10 Hz to approximately $10^3$ Hz,
at which point the neutron stars collide and coalesce. 
The inspiral waveforms are determined to high accuracy by only a few 
parameters: the masses and spin angular momenta of the binary components, 
and the initial orbital elements (i.e., the elements when the waves enter
the detector band). As the binary's bodies spiral closer and closer together,
the waveform increases in amplitude and sweeps up in frequency
(i.e., undergoing a ``chirp''). The shapes of the waves, i.e., the
waves' harmonic content, are determined by the orbital eccentricity.
Gravitational radiation energy losses should lead to highly circular 
binary orbits.
In the Newtonian/quadrupole approximation, for a circular orbit,
the rate at which the frequency sweeps or ``chirps'', $df/dt$, is determined 
solely by the binary's ``intrinsic chirp mass'',
${\cal M}_0 \equiv (M_1 M_2)^{3/5}/(M_1+M_2)^{1/5}$,
where $M_1$ and $M_2$ are the two bodies' masses.
The number of cycles spent near a given frequency is $n=f^2 (df/dt)^{-1}$.
In LIGO, the frequency sweep is monitored by matched filters.
The incoming noisy signal is cross correlated with theoretical templates.
\cite{Thorne95}

For a binary inspiral source located at redshift $z$,
the detectors measure ${\cal M} \equiv {\cal M}_0 (1+z)$, 
which is referred to as the {\it observed} chirp mass. 
For a given detector, the signal-to-noise ratio is \cite{Finn96}
\begin{equation}
\rho= 8 \Theta\, \frac{r_0}{d_L}\left( \frac{ {\cal M}}{1.2 
{\rm M}_{\odot}}\right)
^{5/6} \, \zeta(f_{\rm max}),
\label{eq:rho}
\end{equation}
$d_L$ is our luminosity distance to the binary inspiral source.
$r_0$ and $\zeta(f_{\rm max})$ depend only on the detector's
noise power spectrum.  
The characteristic distance $r_0$ gives an overall sense of the depth
to which the detector can ``see''.
For advanced LIGO, $r_0=355\,$Mpc. \cite{Finn96}
$0 \leq \zeta(f_{\rm max}) \leq 1$ is a dimensionless function,
its argument $f_{\rm max}$ is the redshifted instantaneous orbital
frequency when the inspiral terminates because of the
coalescence (or the imminence of coalescence, when the orbital evolution is
no longer adiabatic) of the compact components;
$\zeta$ reflects the overlap of the signal power with the detector bandwidth.
For source redshift $z$, $\zeta \simeq 1$ for $1+z \leq 10\,[2.8 
{\rm M}_{\odot}/(M_1+M_2)]$ \cite{Finn96}. $\zeta \simeq 1$ is 
a good approximation in the context of this paper.
$\Theta$ is the angular orientation function, it arises from
the dependence of $\rho$ on the relative orientation of the source
and the detector, $0\leq \Theta \leq 4$. Although $\Theta$ can not
be measured, its probability distribution can be approximated by
\cite{FinnCher93}:
\begin{equation}
P_{\Theta}(\Theta)= \left\{ \begin{array}{ll}5\Theta (4-\Theta)^3 /256,
\hskip 1cm &\mbox{if $0<\Theta<4$},
\\0, \hskip 2cm &\mbox{otherwise}. 
\end{array} 
\right.
\label{eq:P(Theta)}
\end{equation}
$P_{\Theta}(\Theta)$ peaks at $\Theta=1$.

Throughout this paper, in presenting our results, we use
$\rho\geq \rho_0=8$, $r_0=355\,$Mpc [for advanced LIGO], 
and ${\cal M}_0=1.2\,{\rm M}_{\odot}$ [for typical neutron star binaries].

\section{Maximum redshift of observed events}

The luminosity distance $d_L(z)=(1+z)^2 d_A(z)$, where $d_A(z)$
is the angular diameter distance. Let us denote the fractions of
the critical density as $\Omega_0$ (matter), $\Omega_{\Lambda}$
(the cosmological constant), and $\Omega_k$ (curvature). Note that $\Omega_0+\Omega_{\Lambda}+\Omega_k=1$.
Then we have
\begin{equation}
\frac{(1+z)\,d_A(z)}{c H_0^{-1}}= \left\{ \begin{array}{lll}
\sin\left[ \sqrt{|\Omega_k|}\, \Gamma(z)\right] /\sqrt{|\Omega_k|}, 
\hskip 1cm  &\mbox{$\Omega_k<0$ (closed)},\\
\Gamma(z), 
\hskip 1cm  &\mbox{$\Omega_k=0$ (flat)},\\
\sinh\left[ \sqrt{\Omega_k}\, \Gamma(z)\right]/\sqrt{\Omega_k}, 
\hskip 1cm  &\mbox{$\Omega_k>0$ (open)}.
\end{array} \right.
\end{equation}
$\Gamma(z)$ is defined to be 
\begin{equation}
\label{eq:Gamma(z)}
\Gamma(z) \equiv \int^z_0 \frac{ {\rm d} w}{\sqrt{ \Omega_0 (1+w)^3+
\Omega_{\Lambda}+ \Omega_k (1+w)^2}}.
\end{equation}
In a flat Universe,
\begin{eqnarray}
\frac{(1+z)\,d_A(z)}{c H_0^{-1}}&=& \int^z_0 \frac{ {\rm d} w}
{ \sqrt{ (1-\Omega_{\Lambda}) (1+w)^3 + \Omega_{\Lambda}}} \\
&=& z+(1-\Omega_{\Lambda})\, \left[-\frac{3}{4}\, z^2+ 
\frac{5-9\Omega_{\Lambda}}{8}\, z^3+ \frac{162\Omega_{\Lambda}-135\Omega_{\Lambda}^2-35}{64}\,z^4
+{\cal O}(z^5) \right].\nonumber
\end{eqnarray}
For $\Omega_{\Lambda}=0$, $(1+z)\,d_A(z)/(c H_0^{-1})=2\left[1-1/\sqrt{1+z}\right]$.
For $\Omega_{\Lambda}=1$, $(1+z)\,d_A(z)/(c H_0^{-1})=z$.

Given the detector threshold in terms of the minimum signal-to-noise ratio 
$\rho_0$, the maximum redshift of the source that the detector can ``see'',
$z_{\rm max}$, is given by Eq.(\ref{eq:rho}) with $\rho=\rho_0$
and $\Theta=4$, i.e., 
\begin{equation}
(1+z_{\rm max})^{7/6} \left[\frac{d_A(z_{\rm max})}{c H_0^{-1}}\right]=hA,
\label{eq:zmax}
\end{equation}
where we have defined
\begin{equation}
A\equiv 0.4733\,\left(\frac{8}{\rho_0}\right)\, \left(\frac{r_0}{355 \,
\mbox{Mpc}}\right)\, \left( \frac{ {\cal M}_0}{1.2 
{\rm M}_{\odot}}\right)^{5/6}.
\label{eq:A}
\end{equation}
Note that $z_{\rm max}$ depends on $h$ only
in the combination $h\, A(r_0,\rho_0, {\cal M}_0)$.
$z_{\rm max}$ increases with $hA$.

In a flat Universe, we can find approximate analytical solution
to Eq.(\ref{eq:zmax}).
For $\Omega_{\Lambda} \geq 0.5$, $z_{\rm max}$ is given by
\begin{equation}
z_{\rm max} \equiv z_{\rm max}^{(1)}\simeq \frac{h A\,
(1+h A)^{-1/6}}{\left.\left[(1+z)\,d_A(z)/(c H_0^{-1})\right]\right|_{z=hA}},
\label{eq:zmax1}
\end{equation}
For $\Omega_{\Lambda}$ close to 0, $z_{\rm max}$ is given by
\begin{equation}
z_{\rm max} \equiv z_{\rm max}^{(2)}\simeq \frac{0.99\,h A}{\eta(hA)}\,
\left[1-\frac{\Omega_{\Lambda}}{1+hA}\left(\frac{1}{\eta(hA)}-1\right)\right],
\hskip 1cm \eta(z)=\frac{2}{z}\left[1-\frac{1}{\sqrt{1+z}}\right].
\end{equation}
Connecting the large and small $\Omega_{\Lambda}$ approximations 
with an arbitrary smoothing function, we have
\begin{equation}
z_{\rm max}=z_{\rm max}^{(1)}\left[ 1-e^{-5\Omega_{\Lambda}^2}\right]
+z_{\rm max}^{(2)}\, e^{-5\Omega_{\Lambda}^2}.
\label{eq:zmaxflat}
\end{equation}
The above approximate formula is accurate to better than 1.2\% for 
$\Omega_{\Lambda}<0.5$ and to better than 0.6\% for $\Omega_{\Lambda}
\geq 0.5$.

\section{Distributions of observed events in redshift and signal-to-noise
ratio}

The number of binary inspiral events seen by a detector on earth 
per source redshift interval per signal-to-noise interval is
\begin{eqnarray}
\frac{{\rm d}\dot{N}(>\rho_0)}{{\rm d} z\, {\rm d} \rho}
&=& \dot{n}\, \cdot 4\pi  \left[d_A(z)\right]^2\, 
\frac{ c\,{\rm d} t}{{\rm d} z}\, P_{\rho}(\rho, z)\nonumber \\
&=&4\pi \dot{n}_0 \left(cH_0^{-1}\right)^3 \, \left[ 
\frac{d_A(z)}{cH_0^{-1}}\right]^2\,
\frac{ 1+z}{ \sqrt{ \Omega_0 (1+z)^3 + 
\Omega_{\Lambda}+ \Omega_k (1+z)^2} }\,P_{\rho}(\rho, z),
\label{eq:P(rho,z)}
\end{eqnarray}
where ${\rm d}t$ is the differential light travel time.
$P_{\rho}(\rho, z)$ is the probability that the detector ``sees''
a source at redshift $z$ with signal-to-noise ratio $\rho>\rho_0$,
\begin{equation}
P_{\rho}(\rho, z)=P_{\Theta}\left(\Theta(\rho)\right)\, 
\left.\frac{\partial\Theta}{\partial\rho}\right|_z
=\frac{\Theta}{\rho}\, P_{\Theta}\left(\Theta(\rho)\right),
\label{eq:P(rho)}
\end{equation}
where $P_{\Theta}(\Theta)$ is given by Eq.(\ref{eq:P(Theta)}).

It is straightforward to find the expected distribution of observed events
in the source redshift $z$ and in the signal-to-noise
ratio $\rho$. The distribution in $z$ is 
\begin{equation}
P(z, >\rho_0)= \frac{{\rm d}\dot{N}(>\rho_0)/{\rm d} z}{\dot{N}(>\rho_0)}=
\frac{4\pi \dot{n}_0 \left(cH_0^{-1}\right)^3 }{\dot{N}(>\rho_0)}
\, \left[ \frac{d_A(z)}{cH_0^{-1}}\right]^2\,
\frac{ 1+z}{ \sqrt{ \Omega_0 (1+z)^3 + \Omega_{\Lambda}+ \Omega_k (1+z)^2} }
\,C_{\Theta}(x),
\end{equation}
where $C_{\Theta}(x)$ is the probability that a given detector detects 
a binary inspiral at redshift $z$ with signal-to-noise ratio greater 
than $\rho_0$, it decreases with $z$. Because of 
Eq.(\ref{eq:P(rho)}), $C_{\Theta}(x)=\int^{\infty}_x {\rm d}\Theta\, 
P_{\Theta}(\Theta)$, hence
\begin{eqnarray}
C_{\Theta}(x) &= &\left\{ \begin{array}{ll}
(1+x)\,(4-x)^4 /256, \hskip 0.5cm  &\mbox{if $0<x<4$},\\
0,	\hskip 1.5cm		&\mbox{otherwise}. \end{array} \right.\\
x &=&\frac{4}{h\, A(r_0, \rho_0, {\cal M}_0)}\,
(1+z)^{7/6}\left[\frac{d_A(z)}{cH_0^{-1}}\right].
\end{eqnarray}
$x$ is the minimum angular orientation function 
in terms of $z$, $\rho_0$ and $r_0$.
Fig.1 shows the distribution of observed events
in the source redshift $z$, for $\Omega_{\Lambda}=0$,
0.5, 0.9, 1, and for $h=0.5$, 0.8, assuming $A=0.4733$ and a flat Universe. 
$P(z, >\rho_0)$ increases with $z$ due to the increase in volume with $z$,
until $C_{\Theta}(x)$ cuts off the growth at $x \simeq 1$. 
Since at small $z$, $C_{\Theta}(x)$ is most sensitive to $h$ 
[the minimum angular orientation function $x$ scales with $h$],
the peak location of the redshift distribution is determined by $h$.
$\Omega_{\Lambda}$ determines the shape of the peak for given $h$.
Note that while the $h$-dependence of the peak location
comes in through the straightforward scaling of the minimum angular orientation 
function $x$ in the combination of $h\,A(r_0,\rho_0,{\cal M}_0)$
[which depends on detector and source properties],
the shape of the peak depends on $\Omega_{\Lambda}$ 
through the angular diameter distance
$d_A(z)$ in a detector and source independent way.

The distribution in the signal-to-noise ratio $\rho$ is
\begin{eqnarray}
P(\rho, >\rho_0)&= &\frac{{\rm d}\dot{N}(>\rho_0)/{\rm d}\rho}
{\dot{N}(>\rho_0)} \nonumber\\
&= &\frac{4\pi \dot{n}_0 \left(cH_0^{-1}\right)^3 
}{\dot{N}(>\rho_0)} \int^{z_{\rm max}}_0 {\rm d}z\,
\left[ \frac{d_A(z)}{cH_0^{-1}}\right]^2\,
\frac{ 1+z}{ \sqrt{ \Omega_0 (1+z)^3 + \Omega_{\Lambda}+ \Omega_k (1+z)^2}
}\, P_{\rho}(\rho, z),
\end{eqnarray}
where $P_{\rho}(\rho, z)$ is given Eq.(\ref{eq:P(rho)}) with $\Theta(\rho,z)$
given by Eq.(\ref{eq:rho}). In a flat and static universe, 
$P(\rho, >\rho_0)\equiv P(\rho, >\rho_0)_0= 3 \rho_0^3/\rho^4$
\cite{Finn96}.
The number density of events for given signal-to-noise ratio $\rho$ decreases
sharply with increasing $\rho$.
Fig.2 shows the distribution of observed events in the signal-to-noise ratio 
$\rho$ relative to the distribution in a flat and static universe, 
$P(\rho, >\rho_0)/P(\rho, >\rho_0)_0$, for $\Omega_{\Lambda}=0$, 
0.5, 0.9, 1, and for $h=0.5$, 0.8, assuming $A=0.4733$ and a flat Universe.
$P(\rho, >\rho_0)/P(\rho, >\rho_0)_0$ also depends on $h$ through
the combination of $h\,A(r_0,\rho_0,{\cal M}_0)$.

The total number of observed events, $\dot{N}(>\rho_0)$,
is found by integrating Eq.(\ref{eq:P(rho,z)}) over $z$ and $\rho$. 
For the local rate density of Eq.(\ref{eq:localRate}), $\dot{N}(>\rho_0)$
is more sensitive to $h$ than to $\Omega_{\Lambda}$. 
However, note that the $h$ dependence comes in only through 
the overall factor of $\dot{n}_0\,\left(cH_0^{-1}\right)^3$,
and in the combination of $h\,A(r_0,\rho_0, {\cal M}_0)$.
Fig.3 shows the total event rate per year
as function of $h$ in a flat Universe, assuming $A=0.4733$ and 
$\dot{n}_0 = 10^{-7} h\, {\rm Mpc}^{-3} {\rm yr}^{-1}$, 
for $\Omega_{\Lambda}=0$ [solid line], 0.5 [short dashed line], 
0.9 [dotted line], 1 [long dashed line].

\section{Accuracy in the measurement of the cosmological constant}

To measure $\Omega_{\Lambda}$ from the chirp mass spectrum of
observed neutron star binary inspiral events, we need to consider 
the spread in the intrinsic chirp mass caused by the spread 
in the neutron star masses.
This effect has been discussed by Ref.{\cite{Finn96}} in detail,
and it is contained in our parameter $A$. A distribution in 
neutron star masses distorts the observed chirp mass spectrum in a somewhat
symmetric manner {\cite{Finn96}}, while the variations in $\Omega_{\Lambda}$
and $\Omega_0$ distorts the observed chirp mass spectrum in an
asymmetric manner. We neglect the spread in the intrinsic chirp mass
(which is likely small) in the context of this paper,
then the observed chirp mass spectrum is the same as $P(z,\rho>\rho_0)$.

In a flat Universe, the accuracy of measurement of 
the cosmological constant is
roughly given by the Poisson noise $1/\sqrt{N}$ [$N$ is the 
total number of observed events], divided by the ``efficiency''
$Q$ with which the change in $P(z,\rho>\rho_0)$ reflects the
change in the cosmological constant. We can estimate $Q$ to be
the ratio of the percentage change in $P(z,\rho>\rho_0)$
to the percentage change in the cosmological constant.
For $hA=0.8\times 0.4733$, 
the accuracy in measuring $\Omega_{\Lambda}$
is about $(6\sqrt{N})^{-1}$, where $N$ is the total number of
observed events.

Without the assumption of a flat Universe, the measurement of
the cosmological constant is complicated by the
degeneracy of different cosmological models which give rise to the
same redshift distribution of observed events.
This is because $P(z,\rho>\rho_0)$ depends on both 
$\Omega_{\Lambda}$ and $\Omega_0$, a given $P(z,\rho>\rho_0)$ 
will correspond to a family of related models
without the assumption of a flat Universe.
This limits the accuracy of $\Omega_{\Lambda}$
which can be measured from the observed $P(z,\rho>\rho_0)$, independent
of the total number of observed events. 

$\Omega_0$ and $\Omega_{\Lambda}$ affect $P(z,\rho>\rho_0)$ mostly
through the comoving distance [see Eq.(\ref{eq:Gamma(z)})].
The larger $z$, the more differently $\Omega_0$ and $\Omega_{\Lambda}$ 
affect $P(z,\rho>\rho_0)$. 
For given $P(z,\rho>\rho_0)$ with known
$hA$, a good estimate of the correlation
between the uncertainties in $\Omega_0$ and $\Omega_{\Lambda}$ is
\begin{equation}
\Delta\Omega_0 = \frac{2+z_{\rm max}}{(1+z_{\rm max})^2}\,
\Delta\Omega_{\Lambda}.
\end{equation}
For a given cosmological model ($\Omega_{\Lambda}$, $\Omega_0$),
its $z_{\rm max}$ is given by Eq.(\ref{eq:zmax}) for given $hA$,
the above equation gives a family of models ($\Omega_{\Lambda}'$, $\Omega_0'$)
which give rise to approximately the same $z_{\rm max}$ and
practically indistinguishable $P(z,\rho>\rho_0)$.
However, when we increase $r_0$ (or increase $A$), $z_{\rm max}$ increases,
it becomes easier to separate models with different
$\Omega_0$ and $\Omega_{\Lambda}$.

\newcounter{bean}
Fig.4 shows $P(z, \rho>\rho_0=8)$ for three cosmological models
\begin{list}%
{(\arabic{bean})}{\usecounter{bean}}
\item{$\Omega_{\Lambda}=0.4$, $\Omega_0=0.6$;}
\item{$\Omega_{\Lambda}=0$, $\Omega_0=0.1$;}
\item{$\Omega_{\Lambda}=1$, $\Omega_0=1.35$.}
\end{list}
These models are practically indistinguishable 
for $r_0=355\,$Mpc ($A=0.4733$), but distinguishable for 
$r_0=568\,$Mpc ($A=0.7573$).
Fig.5 shows the corresponding distributions in signal-to-noise ratio $\rho$.
Note that in principle, it is possible to
distinguish these three models using the additional information
from the distribution in $\rho$. Increasing $A$ by going to a 
larger $r_0$ definitely helps lifting the
uncertainty in the determination of  $\Omega_{\Lambda}$ and $\Omega_0$;
i.e., the measurement of $\Omega_{\Lambda}$ becomes more feasible as
the detector sensitivity is increased.

\setcounter{bean}{0}
Note that the particular combination of $\Omega_{\Lambda}$ and
$\Omega_0$, $\alpha \equiv \Omega_0 (1+z_{max})^2 -
\Omega_{\Lambda} (z_{max}+2)$, is measured easily and accurately.
The three degenerate curves for $r_0=355\,$Mpc in Fig.4 have 
\begin{list}%
{(\arabic{bean})}{\usecounter{bean}}
\item{$\alpha=0.22$, $q_0=-0.1$;}
\item{$\alpha=0.20$, $q_0=0.05$;}
\item{$\alpha=0.25$, $q_0=-0.325$,}
\end{list}
where $q_0=\Omega_0/2 - \Omega_{\Lambda}$ is the deceleration 
parameter. Clearly, $\alpha$ is more accurately measured than
$q_0$; this is because $\alpha$ is similar to $q_0$
[$\alpha(z_{max}=0)= 2 q_0$], but contains
the observed maximum redshift $z_{max}$.
At $r_0=568\,$Mpc, we have $\alpha=0.65$, 0.27, and 1.38 
for the three models respectively.

The larger $\Omega_{\Lambda}$, the smaller the parameter space in
($\Omega_{\Lambda}$, $\Omega_0$) which can imitate the flat model
with $\Omega_{\Lambda}+\Omega_0=1$, hence the more accurately we
can determine $\Omega_{\Lambda}$.

\section{Conclusion and discussion}

In summary, we have calculated the expected maximum source redshift
$z_{\rm max}$, the source redshift distribution $P(z,>\rho_0)$, the signal-to-noise ratio distribution $P(\rho,>\rho_0)$,
and the total number of events per year $\dot{N}(\rho>\rho_0)$,
for advanced LIGO detectors in a Universe with nonzero cosmological
constant. $z_{\rm max}$, $P(z,>\rho_0)$, and $P(\rho,>\rho_0)$
all depend on $\Omega_{\Lambda}$ and $\Omega_0$ 
in a fundamental way through the
angular diameter distance, and they all depend on $h$ through the
combination $h\,A(r_0,\rho_0, {\cal M}_0)$.
$\dot{N}(\rho>\rho_0)$ is very sensitive to the local binary merger rate
$\dot{n}_0$ through $\dot{n}_0\,\left(cH_0^{-1}\right)^3$, 
the value of which is quite uncertain at this time.

The expected redshift distribution of 
observed events in an advanced LIGO detector
has a robust and sizable dependence 
on the cosmological constant.
Although the redshift distribution has an apparent dependence on $h$
which is more dominant, this dependence on $h$ is superficial in
the sense that it always appears in the combination of $h\, A(r_0,\rho_0,
{\cal M}_0)$ [$A$ is given by Eq.(\ref{eq:A})];
increasing $h$ has exactly the same effect on the redshift distribution
as increasing $r_0$ or ${\cal M}_0^{5/6}$, or decreasing $\rho_0$, by the
same amount. On the other hand, the redshift distribution depends on
$\Omega_{\Lambda}$ and $\Omega_0$ in a fundamental way, this dependence is 
detector and source independent. 

If we live in a flat Universe, then the cosmological constant can
be determined quite accurately from the expected redshift distribution 
of observed events with a cut on the signal-to-noise ratio.
Assuming arbitrary geometry of the Universe, the expected redshift distribution 
of observed events with a cut on the signal-to-noise ratio
may correspond to a family of related cosmological models with
different values of $\Omega_{\Lambda}$ and $\Omega_0$
[the combination $\alpha \equiv \Omega_0 (1+z_{max})^2 -
\Omega_{\Lambda} (z_{max}+2)$ is easily and accurately measured];
this degeneracy can be lifted by either accurately measuring
the distribution of observed events in signal-to-noise ratio, or 
by increasing the detector sensitivity.
However, it may well prove more practical to use other astrophysical 
constraints on $\Omega_{\Lambda}$ and $\Omega_0$ to resolve this degeneracy.
The data from advanced LIGO detectors
should provide an independent and robust measurement of the cosmological 
constant.

\acknowledgments
Y.W. is supported by the DOE and NASA under Grant NAG5-2788.
E.L.T. gratefully acknowledges support from NSF grant AST94-19400.
We thank L.S. Finn for a useful conversation.


\narrowtext

\newpage
\nonfrenchspacing
\parindent=20pt
\centerline{{\bf Figure Captions}}

Fig.1 The distribution of observed events
in the source redshift $z$, for $\Omega_{\Lambda}=0$ 
[solid line], 0.5 [short dashed line], 0.9 [dotted line], 1 [long dashed line], 
and for $h=0.5$, 0.8, assuming $A=0.4733$ and a flat Universe.

Fig.2 The distribution of observed events
in the signal-to-noise ratio $\rho$ relative to the distribution 
in a flat and static universe, for $\Omega_{\Lambda}=0$ 
[solid line], 0.5 [short dashed line], 0.9 [dotted line], 1 
[long dashed line], and for $h=0.5$, 0.8, assuming $A=0.4733$ 
and a flat Universe.

Fig.3 The total event rate per year
as function of $h$ in a flat Universe, assuming $A=0.4733$ and 
$\dot{n}_0 = 10^{-7} h\, {\rm Mpc}^{-3} {\rm yr}^{-1}$, 
for $\Omega_{\Lambda}=0$ [solid line], 0.5 [short dashed line], 
0.9 [dotted line], 1 [long dashed line].

Fig.4 The distribution of observed events in the source redshift $z$,
with $r_0=355\,$Mpc, 568$\,$Mpc ($A=0.4733, 0.7573$),
for three cosmological models:
(1) solid line: $\Omega_{\Lambda}=0.4$, $\Omega_0=0.6$;
(2) dashed line: $\Omega_{\Lambda}=0$, $\Omega_0=0.1$;
(3) dotted line: $\Omega_{\Lambda}=1$, $\Omega_0=1.35$.

Fig.5 The distribution of observed events
in the signal-to-noise ratio $\rho$ relative to the distribution 
in a flat and static universe, for the models in Fig.4,
with the same line types.

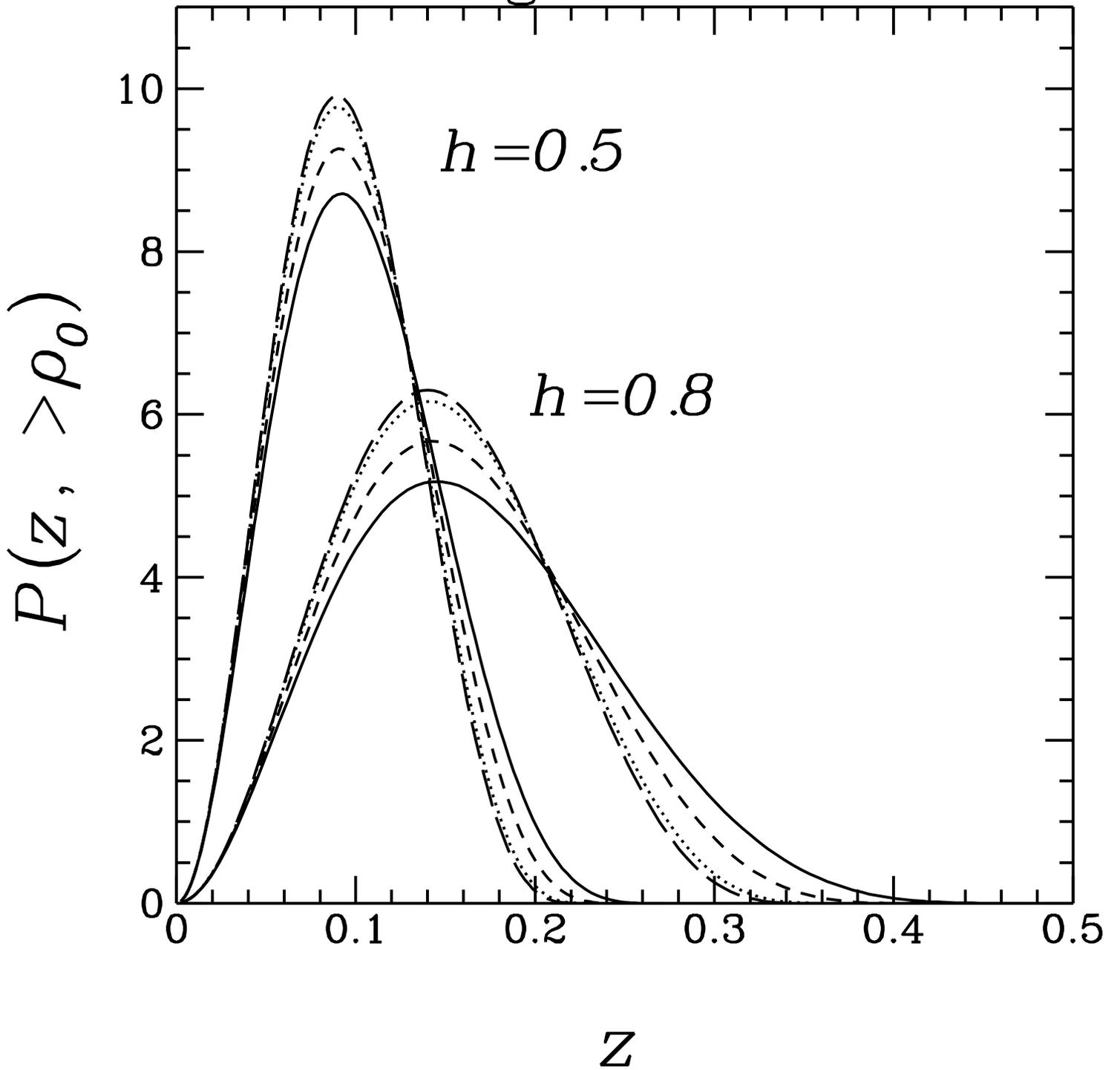
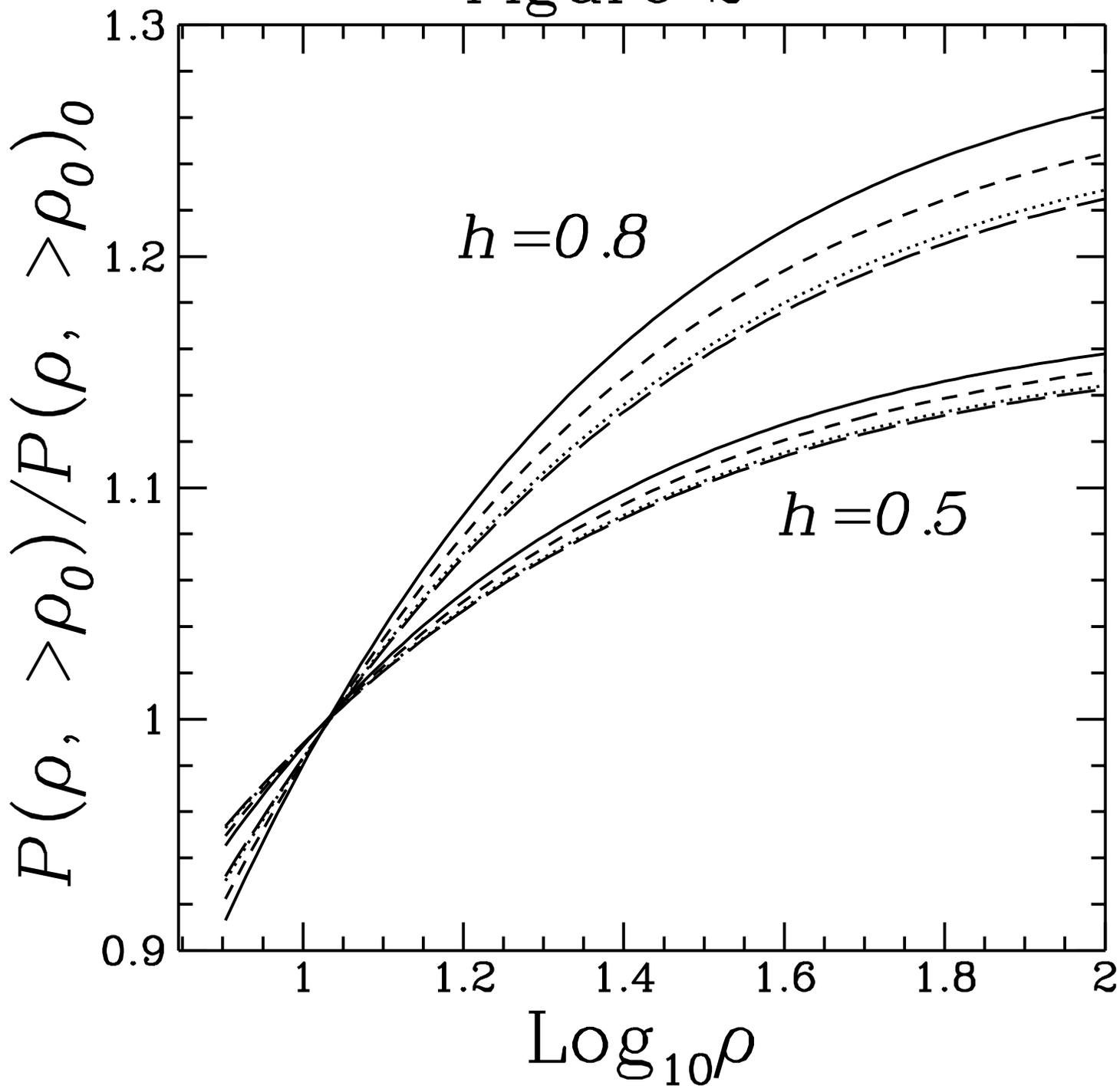
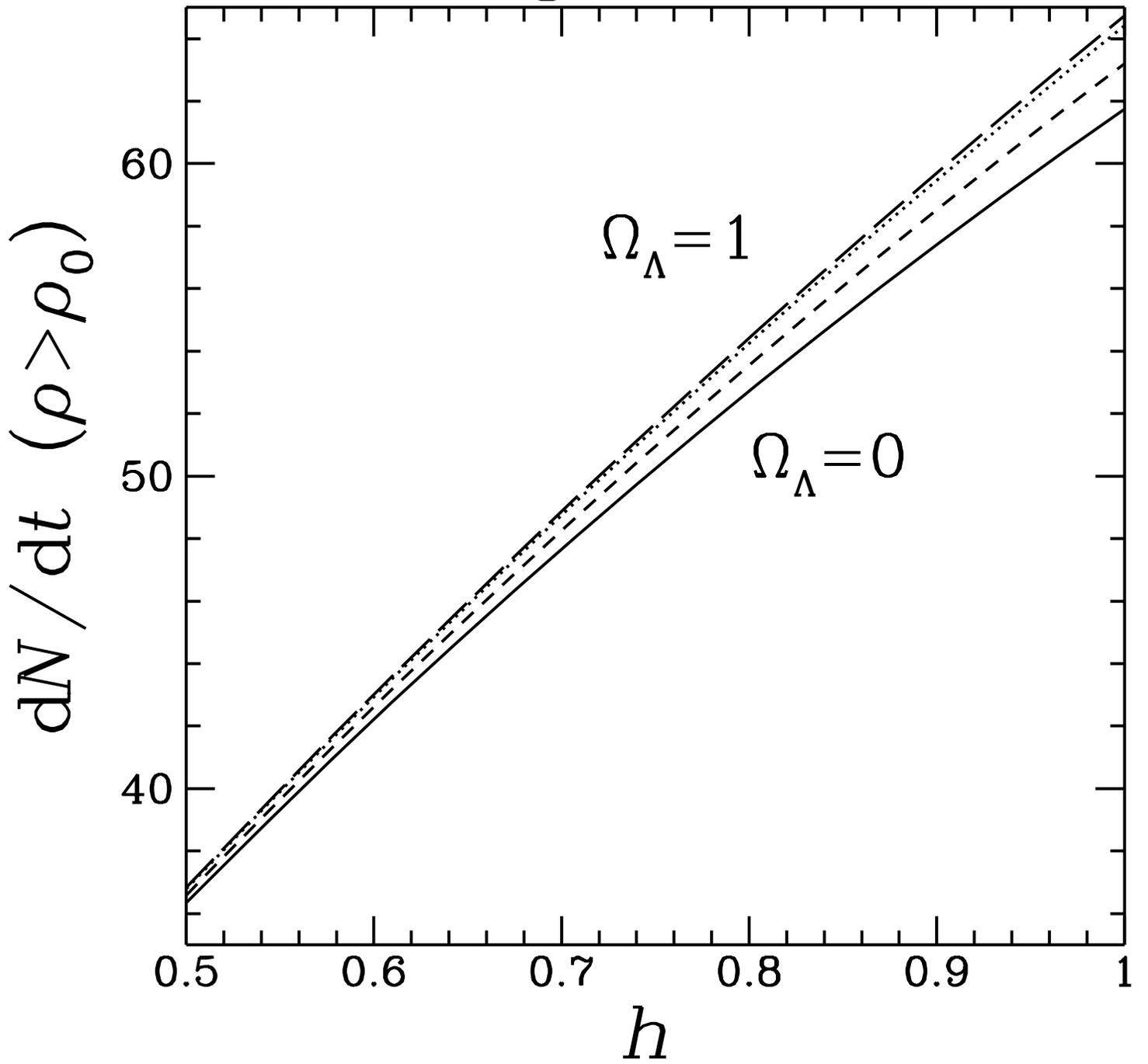
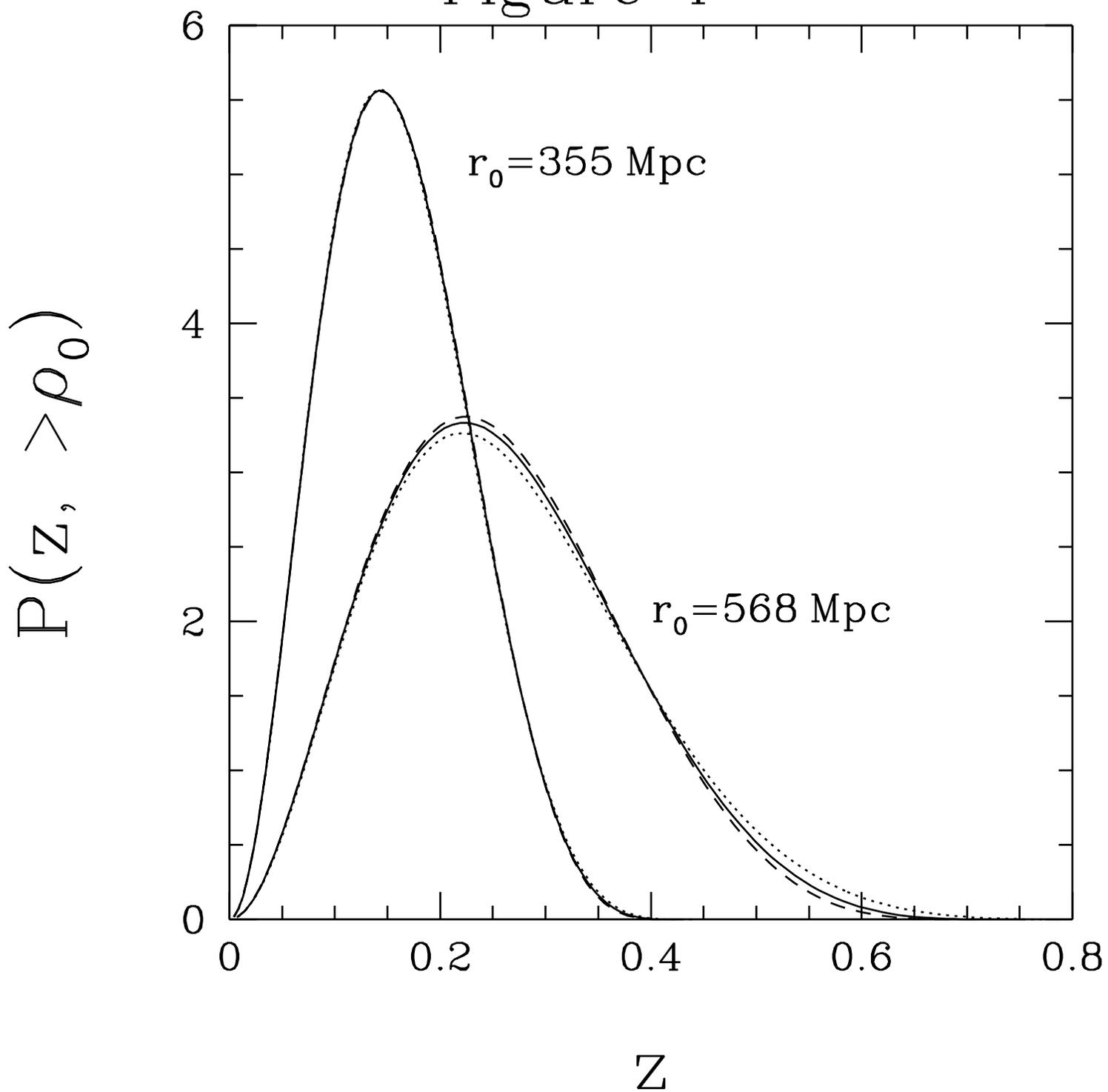
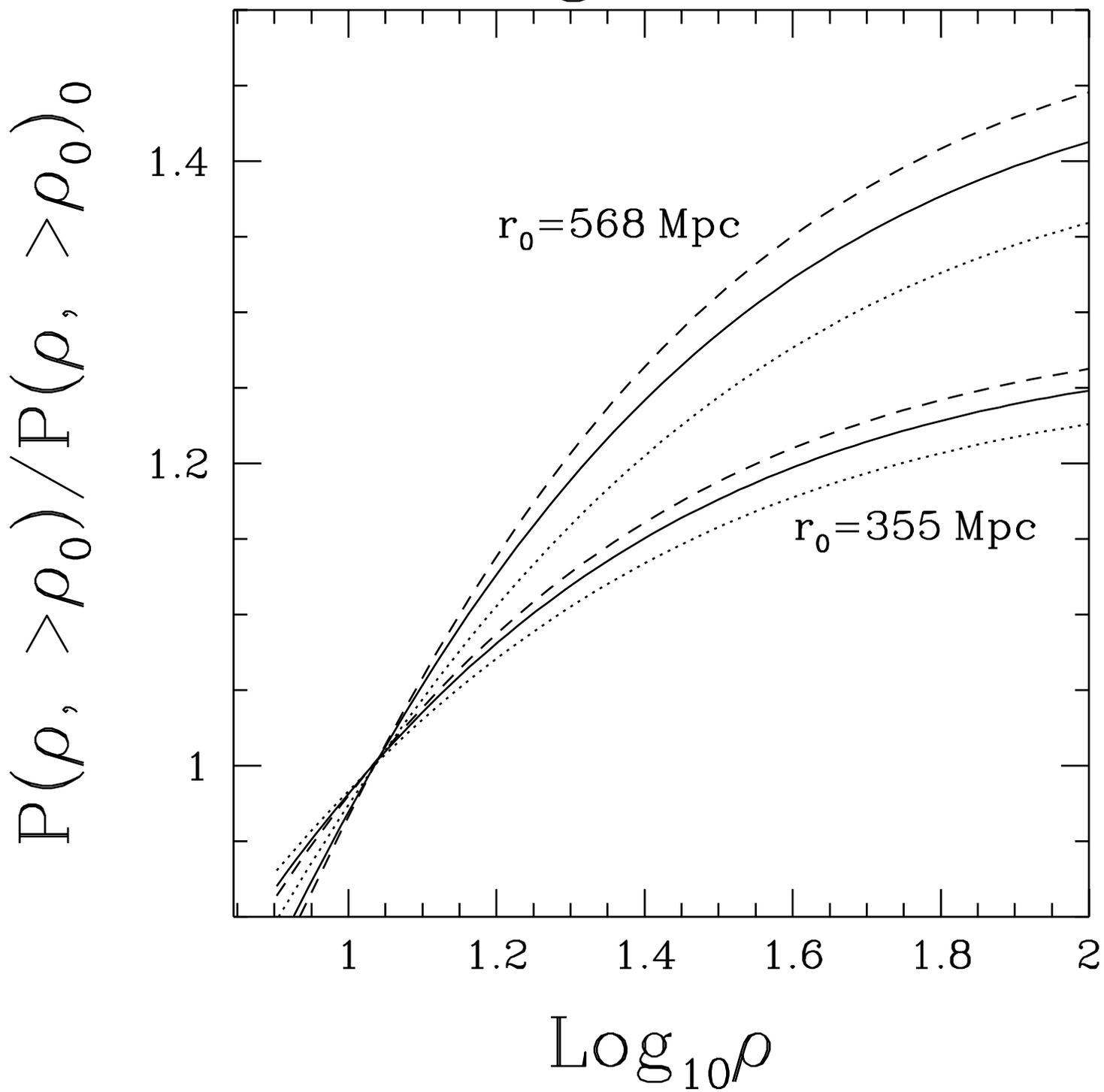
\end{document}